\documentclass[epj]{svjour}
\usepackage{graphics}
\usepackage{graphicx}

\begin{document}
\titlerunning{Noncommutative Rotating Black Hole}
\title{Noncommutative Geometry Inspired Rotating Black Hole in Three Dimensions}

\author{Juan Manuel Tejeiro
\thanks{\emph{email:} jmtejeiros@unal.edu.co}
 \and Alexis Larranaga
\thanks{\emph{email:} ealarranaga@unal.edu.co}%
}                     
\offprints{}          
\institute{Observatorio Astronomico Nacional. Facultad de Ciencias. Universidad
Nacional de Colombia}
\date{Received: date / Revised version: date}
%
\abstract{
We find a new rotating black hole in three-dimensional anti-de Sitter
space using an anisotropic perfect fluid inspired by the noncommutative
black hole. We deduce the thermodynamical quantities of this black
hole and compare them with those of a rotating BTZ solution.
} 
\maketitle
\section{Introduction}
\label{intro}
The theoretical discovery of radiating black holes \cite{hawking}
disclosed the first physically relevant window on the mysteries of
quantum gravity. After many years of intensive research in this field
various aspects of the problem still remain under debate. For instance,
a fully satisfactory description of the late stage of black hole evaporation
is still missing. The string/black hole correspondence principle \cite{sisskind}
suggests that in this extreme regime stringy effects cannot be neglected.
This is just one of many examples of how the development of string
theory has affected various aspects of theoretical physics. Among
different outcomes of string theory, we focus on the result that target
spacetime coordinates become noncommuting operators on a D-brane \cite{witten}.
Thus, string-brane coupling has put in evidence the necessity of spacetime
quantization. 

The noncommutativity of spacetime can be encoded in the commutator

\begin{equation}
\left[x^{\mu},x^{\nu}\right]=i\theta^{\mu\nu},\end{equation}
 where $\theta^{\mu\nu}$ is an anti-symmetric matrix which determines
the fundamental cell discretization of spacetime much in the same
way as the Planck constant $\hbar$ discretizes the phase space. This
noncommutativity provides a black hole with a minimum scale $\sqrt{\theta}$
known as the noncommutative black hole \cite{key-1,key-2,key-3,key-4,key-5,key-7},
whose commutative limit is the Schwarz-schild metric. Myung and Kim,
\cite{myung1} have studied the thermodynamics and evaporation process
of this noncommutative black hole while the entropy issue of this
black hole was discussed in \cite{banarjee1,banarjee2} and Hawking
radiation was considered in \cite{nozari}. 

In this paper, we construct a new rotating black hole in AdS3 spacetime
using an anisotropic perfect fluid inspired by the 4D noncommutative
black hole, resulting in a solution with two horizons. We compare
the thermodynamics of this black hole with that of a rotating BTZ
black hole \cite{BTZ1,BTZ2}.

\section{Derivation of the Rotating Solution}

It has been shown \cite{key-1,key-2} that the noncommutativity eliminates
point-like structures in favor of smeared objects in flat space-time.
A way of implementing the effect of smearing is a substitution rule:
in four-dimensional (4D) space-times, the Dirac delta function $\delta^{4D}\left(r\right)$
is replaced by a Gaussian distribution with minimal width $\sqrt{\theta}$
\cite{key-1,key-2,key-3,key-4,key-5,key-7},

\begin{equation}
\rho^{4D}\left(r\right)=\frac{M}{\left(4\pi\theta\right)^{3/2}}e^{-r^{2}/4\theta}\end{equation}

and the corresponding mass distribution is given by

\begin{equation}
m^{4D}\left(r\right)=4\pi\int_{0}^{r}r'^{2}\rho^{4D}\left(r'\right)dr'=\frac{2M}{\sqrt{\pi}}\gamma\left(\frac{3}{2},\frac{r^{2}}{4\theta}\right),\end{equation}
where $\gamma\left(\frac{3}{2},\frac{r^{2}}{4\theta}\right)$ is the
lower incomplete gamma function defined as

\begin{equation}
\gamma\left(a,z\right)=\int_{0}^{z}t^{a-1}e^{-t}dt.\end{equation}

In three dimensions, the Dirac delta function $\delta^{3D}\left(r\right)$
is replaced by a Gaussian distribution with minimal width $\sqrt{\theta}$,
\cite{myung2},

\begin{equation}
\rho^{3D}\left(r\right)=\frac{M}{4\pi\theta}e^{-r^{2}/4\theta}\end{equation}

and the corresponding mass distribution is now

\begin{eqnarray}
m^{3D}\left(r\right) & = & 2\pi\int_{0}^{r}r'\rho^{3D}\left(r'\right)dr'=M\gamma\left(1,\frac{r^{2}}{4\theta}\right)\\
 & = & M\left(1-e^{-r^{2}/4\theta}\right).\end{eqnarray}

In order to find a black hole solution in $AdS_{3}$ space-time, we
introduce the Einstein equation

\begin{equation}
R_{\mu\nu}-\frac{1}{2}g_{\mu\nu}R=8\pi T_{\mu\nu}+\frac{1}{\ell^{2}}g_{\mu\nu}\end{equation}
where $\ell$ is related with the cosmological constant by 

\begin{equation}
\Lambda=-\frac{1}{\ell^{2}}.\end{equation}
The energy-momentum tensor will take the anisotropic form

\begin{equation}
T_{\nu}^{\mu}=\mbox{diag}\left(-\rho,p_{r},p_{\bot}\right).\end{equation}

In order to completely define this tensor, we rely on the covariant
conservation condition $T_{\quad,\nu}^{\mu\nu}=0$. This gives the
source as an anisotropic fluid of density $\rho$, radial pressure

\begin{equation}
p_{r}=-\rho\end{equation}

and tangential pressure

\begin{equation}
p_{\bot}=-\rho-r\partial_{r}\rho.\end{equation}

Solving the above equations, we find the line element

\begin{equation}
ds^{2}=-f\left(r\right)dt^{2}+f^{-1}\left(r\right)dr^{2}+r^{2}\left(d\varphi+N^{\varphi}dt\right)^{2},\label{eq:solucionBTZ}\end{equation}

where 

\begin{eqnarray}
f\left(r\right) & = & -8M\left(1-e^{-r^{2}/4\theta}\right)+\frac{r^{2}}{l^{2}}+\frac{J^{2}}{4r^{2}}\\
N^{\varphi} & = & -\frac{J}{2r^{2}}.\end{eqnarray}

Note that when $\frac{r^{2}}{4\theta}\rightarrow\infty$, either for
considering a large black hole $\left(r\rightarrow\infty\right)$
or for considering the commutative limit $\left(\theta\rightarrow0\right)$,
we obtain the well known BTZ rotating solution with angular momentum
$J$ and total mass $M$,

\begin{equation}
f^{BTZ}\left(r\right)=-8M+\frac{r^{2}}{l^{2}}+\frac{J^{2}}{4r^{2}}.\end{equation}

The line element (\ref{eq:solucionBTZ}) describes the geometry of
a noncommutative black hole with event horizons given by the condition

\begin{equation}
f\left(r_{\pm}\right)=-8M\left(1-e^{-r_{\pm}^{2}/4\theta}\right)+\frac{r_{\pm}^{2}}{l^{2}}+\frac{J^{2}}{4r_{\pm}^{2}}=0.\end{equation}
This equation cannot be solved in closed form. However, by plotting
$f\left(r\right)$ one can read intersections with the $r$-axis and
determine numerically the existence of horizon(s) and their radii.
Fig. 1 shows that the existence of angular momentum introduces new
behavior with respect to the noncommutative black hole studied by
Myung and Yoon \cite{myung2} and others reported before \cite{otro1,otro2}.
Instead of a single event horizon, there are different possibilities:

1. Two distinct horizons for $M>M_{o}$

2. One degenerate horizon (extremal black hole) for $M=M_{o}$

3. No horizon for $M<M_{o}$.

\begin{figure}
\resizebox{0.5\textwidth}{!}{%
  \includegraphics{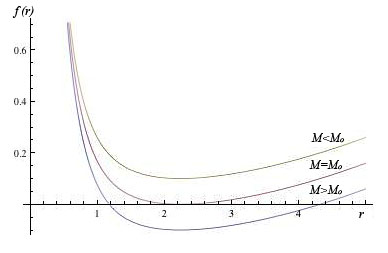}
}
\caption{Metric function $f$ as a function of $r$.
We have taken the values $\theta=0.1$, $\ell=10$ and $J=1$. The
minimum mass is $M_{o}\approx0.0125$.}
\label{fig:1}       
\end{figure}

In view of this results, there can be no black hole if the original
mass is less than the lower limit mass $M_{o}$. The horizon of the
extremal black hole is determined by the conditions $f=0$ and $\partial_{r}f=0,$
which gives

\begin{equation}
r_{o}^{4}\left[\frac{1-\left(1+\frac{r_{o}^{2}}{4\theta}\right)e^{-r_{o}^{2}/4\theta}}{1-\left(1-\frac{r_{o}^{2}}{4\theta}\right)e^{-r_{o}^{2}/4\theta}}\right]=\frac{J^{2}\ell^{2}}{4}\end{equation}
and then, the mass of the extremal black hole can be written as 

\begin{equation}
M_{o}=\frac{\left(\frac{r_{o}^{2}}{\ell^{2}}+\frac{J^{2}}{4r_{o}^{2}}\right)}{8\left(1-e^{-r_{o}^{2}/4\theta}\right)}.\end{equation}

In the commutative limit, $\theta\rightarrow0$, the extreme black
hole has the horizon at

\begin{equation}
r_{o}^{BTZ}=\sqrt{\frac{J\ell}{2}}\end{equation}
and its mass is

\[
M_{o}^{BTZ}=\frac{1}{8}\frac{J}{\ell}.\]

\section{Thermodynamics}

The black hole temperature is given by

\begin{equation}
T_{H}=\frac{1}{4\pi}\left.\partial_{r}f\right|_{r_{+}}\end{equation}

\begin{equation}
T_{H}=\frac{r_{+}}{2\pi\ell^{2}}\left[1-\frac{J^{2}\ell^{2}}{4r_{+}^{4}}+\frac{\left(r_{+}^{2}+\frac{J^{2}\ell^{2}}{4r_{+}^{2}}\right)}{4\theta\left(1-e^{r_{+}^{2}/4\theta}\right)}\right].\end{equation}

For large black holes, i.e. $\frac{r_{+}^{2}}{4\theta}>>0$, one recovers
the temperature of the rotating BTZ black hole,

\begin{equation}
T_{H}^{BTZ}=\frac{r_{+}}{2\pi\ell^{2}}\left[1-\frac{J^{2}\ell^{2}}{4r_{+}^{4}}\right].\end{equation}

As shown in the Fig. 2., the temperature is a monotonically increasing
function of the horizon radius for large black holes and the temperature
of the extreme black hole is zero.

\begin{figure}
\resizebox{0.5\textwidth}{!}{%
  \includegraphics{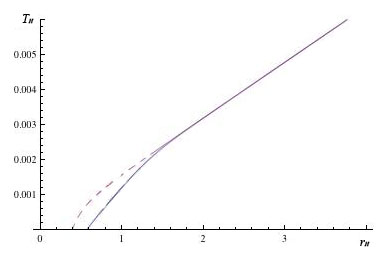}
}
\caption{Hawking temperature versus $r_{H}$. The
solid line represents the temperature for the noncommutative black
hole with $\theta=0.1$. The dashed line represents the temperature
for the rotating BTZ black hole. In both cases we have taken the values
$\ell=10$ and $J=0.03$. }
\label{fig:2}       
\end{figure}

{
The first law of thermodynamics for a rotating black hole reads

\begin{equation}
dM=T_{H}dS+\Omega dJ,\end{equation}

where the angular velocity of the black hole is given by

\begin{equation}
\Omega=\left(\frac{\partial M}{\partial J}\right)_{r_{+}}=\frac{J}{2r_{+}^{2}},\end{equation}
that is exactly the same of the rotating BTZ solution. We calculate
the entropy as

\begin{equation}
S=\int_{r_{o}}^{r_{+}}\frac{1}{T_{H}}dM\end{equation}

which finally gives

\begin{equation}
S=\frac{\pi}{2}\int_{r_{o}}^{r_{+}}\left(\frac{1}{1-e^{-\xi^{2}/4\theta}}\right)d\xi.\end{equation}

The entropy as a function of $r_{+}$ is depicted in Fig. 3. Note
that, in the large black hole limit, the entropy function corresponds
to the Bekenstein-Hawking entropy (area law), $S_{BH}=\frac{\pi r_{+}}{2}$,
for the rotating BTZ geometry.

\begin{figure}
\resizebox{0.5\textwidth}{!}{%
  \includegraphics{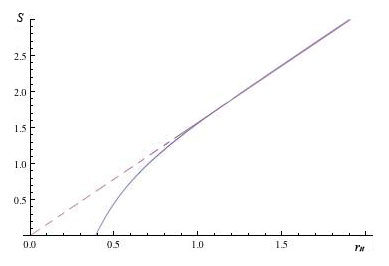}
}
\caption{Entropy versus $r_{H}$. The solid line represents
the entropy of the noncommutative black hole with $\theta=0.1$. The
dashed line represents the entropy of the rotating BTZ black hole.
In both cases we use $\ell=10$. }
\label{fig:3}       
\end{figure}

\section{Conclusion}

We construct a noncommutative rotating black hole in $AdS_{3}$ spacetime
using an anisotropic perfect fluid inspired by the 4D noncommutative
black hole. As well as its $4D$ counterpart, this black hole has
two horizons. We compare the thermodynamics of this black hole with
that of a rotating BTZ black hole. The Hawking temperature, angular
velocity and entropy of large noncommutative black hole approach those
of BTZ.

\end{document}